\documentclass{article}

\usepackage{PRIMEarxiv}
\usepackage{amssymb}
\usepackage[T1]{fontenc}
\usepackage{graphicx}
\usepackage{color}
\graphicspath{{images/}} 
\usepackage{color, soul}
\usepackage{microtype}
\soulregister\cite7
\soulregister\ref7
\soulregister\pageref7
\usepackage{multirow}
\usepackage{comment}
\usepackage{amssymb}
\usepackage{pifont}
\usepackage{amsmath} 
\usepackage[linesnumbered,ruled,vlined]{algorithm2e}
\usepackage{caption}
\usepackage{subcaption}
\usepackage{dsfont}
\usepackage{url}
\usepackage{tikz}
\usetikzlibrary{shapes.geometric, arrows}
\tikzstyle{box} = [rectangle, rounded corners, minimum width=0.2\linewidth, minimum height=1cm,text centered, text width=0.2\linewidth, draw=black, fill=white!30]

\tikzstyle{arrow} = [thick,->,>=stealth]

\title{Protecting Deep Neural Network Intellectual Property with Chaos-Based White-Box Watermarking}

\author{
Sangeeth B.\\
Department of Computer Applications, \\
Cochin University of Science \\
and Technology, India \\
\texttt{sangeeth@cusat.ac.in} \\
\And
Serena Nicolazzo \\
Department of Science and Technological Innovation, \\
University of Eastern Piedmont, \\
V.le Teresa Michel, 11, Alessandria, Italy\\
\texttt{serena.nicolazzo@uniupo.it} \\
\And    
Deepa K.\\
Department of Computer Science and Engineering,\\
SCMS School of Engineering and Technology,\\
Ernakulam, Kerala, India \\
\texttt{deepak@scmsgroup.org} \\
\And
Vinod P. \\
Department of Computer Applications, \\
Cochin University of Science \\
and Technology, India \\
\texttt{vinod.p@cusat.ac.in} \\
}

\begin{document}

\maketitle

\date{September 2025}

\begin{abstract}
The rapid proliferation of deep neural networks (DNNs) across commercial, academic, and mission-critical domains has led to increasing concerns regarding intellectual property (IP) protection and model misuse. Trained DNNs represent valuable assets, often developed through significant investments in data, computation, and expertise. However, the ease with which models can be copied, redistributed, or repurposed highlights the urgent need for effective mechanisms to assert and verify model ownership. In this work, we propose an efficient and resilient white-box watermarking framework that embeds ownership information into the internal parameters of a DNN using chaotic sequences. The watermark is generated using a logistic map, a well-known chaotic function, producing a sequence that is sensitive to its initialization parameters. This sequence is injected into the weights of a chosen intermediate layer without requiring structural modifications to the model or degradation in predictive performance. To validate ownership, we introduce a verification process based on a genetic algorithm that recovers the original chaotic parameters by optimizing the similarity between the extracted and regenerated sequences. 
The effectiveness of the proposed approach is demonstrated through extensive experiments on image classification tasks using MNIST and CIFAR-10 datasets. The results show that the embedded watermark remains detectable after fine-tuning, with negligible loss in model accuracy. In addition to numerical recovery of the watermark, we perform visual analyses using weight density plots and construct activation-based classifiers to distinguish between original, watermarked, and tampered models. These auxiliary techniques enhance the interpretability and reliability of watermark detection. Overall, the proposed method offers a flexible and scalable solution for embedding and verifying model ownership in white-box settings. Its compatibility with different stages of the model lifecycle, either during training or after deployment, makes it well-suited for real-world scenarios where IP protection is critical.
\end{abstract}

\keywords{Watermarking, Deep Neural Network, Chaotic sequence, Logistic regression classifier, Genetic Algorithm, Watermark verification}

\section{Introduction}

In recent years, Deep Neural Networks (DNNs) have become integral to the development of intelligent systems across a wide range of domains, including computer vision, natural language processing, healthcare diagnostics, autonomous systems, and finance. These models achieve exceptional performance due to their ability to learn hierarchical representations from large-scale datasets. However, training such models typically requires substantial investment in terms of data acquisition, computational resources, architecture design, and hyperparameter tuning. As a result, trained DNNs have become valuable Intellectual Property (IP), and protecting them from unauthorized use, theft, or redistribution has emerged as a pressing concern within the machine learning community \cite{xue2021intellectual}. Unlike traditional software, where licensing and access controls can be more easily enforced, neural networks are typically stored and shared as parameterized weight files or serialized model graphs. This makes it easy for malicious actors to copy, fine-tune, or repurpose models without detection or attribution. Once a model is trained and publicly released or shared across platforms, there is a risk that the ownership of that model may be compromised. To address this challenge, researchers have begun exploring digital watermarking as a means to assert and verify model ownership.

Watermarking is a well-established concept in the field of digital media, where imperceptible data is embedded into content such as images, audio, or video to prove authorship, track distribution, or detect tampering \cite{cox2002digital}. In the context of deep learning, watermarking refers to the process of embedding ownership information into the model itself, in a way that allows for subsequent verification, ideally without affecting the model's original performance. The embedded watermark serves as a hidden signature that can be extracted or verified by the rightful owner to confirm the legitimacy of a model \cite{singh2023comprehensive}.

Deep Neural Network watermarking can be approached from multiple angles, depending on the assumptions made about the level of access to the model and the method used for embedding the watermark. One of the primary distinctions in watermarking approaches is between white-box and black-box methods. White-box watermarking assumes that the internal parameters of the model, such as weights or feature maps, are accessible during verification. In such cases, the watermark can be embedded directly into specific layers or weight tensors, and verified by inspecting their structure or behavior. On the other hand, black-box watermarking assumes only query access to the model. Verification is performed by observing the model's outputs in response to a set of specially crafted inputs, known as triggers. While black-box methods are more applicable to deployment scenarios such as Machine Learning-as-a-Service (MLaaS) platforms, they tend to be less robust and more vulnerable to output perturbation or trigger inversion. In addition to the access paradigm, watermarking techniques can be categorized based on how the watermark is embedded. Static watermarking methods embed information into the model's internal structure, such as by altering the distribution of weights or activations. These methods often rely on additional loss functions or regularization terms during training to guide the embedding process. For example, some approaches introduce loss terms that force a subset of weights to follow a particular pattern or encode specific bit sequences. Other methods manipulate activation patterns in response to selected inputs to encode the watermark. Dynamic watermarking, in contrast, embeds the watermark into the model's behavior. This is typically achieved by injecting adversarial examples or synthetic inputs that cause the model to produce predefined outputs when the watermark is present. Dynamic methods are commonly used in black-box settings and often rely on backdoor-style mechanisms. However, such techniques may inadvertently degrade model generalization or raise ethical concerns, especially when used in safety-critical applications. 

An effective watermarking method must satisfy several important criteria. First, it must ensure fidelity, meaning that the performance of the watermarked model on its primary task remains unaffected. The presence of the watermark should not lead to a reduction in accuracy or introduce unintended outputs. Second, the watermark must be robust to common post-processing operations such as fine-tuning, pruning, quantization, and model compression. These transformations are frequently applied in practice to adapt or optimize pre-trained models, and the watermark should remain intact through such modifications. Third, the watermark should be secure and stealthy. It must not be easily detectable, removable, or forgeable by an adversary, even if they have some knowledge of the embedding process or model structure. Ideally, the watermark should also be verifiable without requiring access to the training data or extensive retraining.

Despite a growing body of work on DNN watermarking, several challenges remain unresolved. Many existing techniques require modifying the training pipeline or introducing additional loss components, which may not be feasible for third-party or legacy models. Some approaches are fragile and can be easily broken by adversarial fine-tuning or surrogate model attacks. Others introduce noticeable alterations to model behavior or accuracy, reducing their practical utility. Furthermore, most methods are designed for image classification tasks and convolutional architectures, and their applicability to other domains, such as natural language processing, audio recognition, or graph learning, has not been extensively validated \cite{barni2021dnn}.

Another major limitation in current watermarking schemes is the reliance on handcrafted or static watermark signals. These signals, once identified, can be filtered or removed through parameter smoothing or adversarial training. To counter this, researchers have begun exploring more adaptive and unpredictable watermark generation strategies. One such strategy involves leveraging chaotic systems to generate watermark sequences. Chaotic sequences possess high sensitivity to initial conditions and are difficult to replicate without precise knowledge of the generating parameters. This makes them well-suited for watermarking applications where uniqueness and resilience are critical. Chaotic watermarking involves embedding a signal derived from a deterministic but unpredictable function, such as the logistic map, into the weights of a neural network. The resulting watermark is both deterministic, allowing for recovery through parameter search, and highly sensitive, making it difficult to forge. Recovering the watermark from a model involves searching for the original chaotic parameters that produce a sequence closely matching the embedded watermark. This recovery task is non-trivial due to the nonlinear nature of the chaotic system and the noise introduced by model transformations. As such, heuristic optimization methods such as Genetic Algorithms (GAs) have proven effective for verifying chaotic watermarks. GAs simulate natural selection processes and are capable of efficiently searching large, complex spaces, making them ideal for this task.

In this paper, we propose a white-box watermarking technique that uses chaotic sequences for watermark generation and genetic algorithms for ownership verification. The watermark is embedded by adding a scaled chaotic sequence to the weights of a selected intermediate layer in a pre-trained model. The modification is kept minimal to preserve model accuracy. After embedding, the model is fine-tuned on the original training data to further mitigate any performance impact. For verification, the watermark signal is extracted from the weight differences and compared against regenerated chaotic sequences. A genetic algorithm is used to optimize the chaotic parameters by minimizing the distance between the extracted and regenerated sequences, ensuring that the watermark can be reliably recovered even under weight perturbations.

The main contributions of this work are summarized as follows:
\begin{itemize}
    \item We propose a lightweight white-box watermarking technique that leverages chaotic sequences for watermark generation and embeds the watermark by minimally perturbing the weights of a selected intermediate layer in a pre-trained deep neural network. This approach ensures that the watermarking process does not require altering the original training procedure or retraining the model from scratch.
    \item To preserve the model's predictive performance, the watermarked model is fine-tuned on the original training data, allowing any minor performance degradation introduced by the watermark to be mitigated effectively.
    \item We introduce a genetic algorithm-based optimization framework for ownership verification, that searches for the optimal chaotic parameters by minimizing the distance between the extracted and regenerated watermark signals. This evolutionary approach enables robust watermark recovery even when the model undergoes weight perturbations due to fine-tuning or noise.
    \item We conduct a comprehensive experimental evaluation on standard image classification benchmarks, including MNIST and CIFAR-10, to demonstrate the fidelity, detectability, and robustness of the proposed method under various conditions.
    \item In addition to quantitative verification, we incorporate auxiliary detection mechanisms, such as activation-based classification analysis and weight density visualization, to further support the presence of the watermark. These components provide a multi-layered verification strategy, enhancing the reliability of ownership claims.
\end{itemize}

Overall, this work contributes a practical and effective approach to white-box watermarking, combining the unpredictability of chaos with the search capabilities of evolutionary algorithms. 

The remainder of this paper is organized as follows. Section \ref{sec:related} presents a comprehensive review of related work on DNN watermarking techniques, highlighting key developments and limitations. Section \ref{sec:water} introduces the architecture of the proposed watermarking framework, detailing the embedding and verification phases based on chaotic sequences and genetic algorithms. Section \ref{sec:experiment} describes the experimental setup and datasets used to evaluate the effectiveness and robustness of the proposed method. Section \ref{sec:discussion} discusses the strengths and weaknesses of the approach and outlines the limitations observed. Finally, Section \ref{sec:conclusion} concludes the paper with a summary of findings and potential directions for future research.

\section{Related Works}
\label{sec:related}

With the increasing commercialization and deployment of Deep Neural Networks (DNNs), protecting their intellectual property has become a significant research concern. A variety of watermarking techniques have been proposed to address unauthorized use, distribution, or ownership disputes of trained DNNs. These methods are generally classified based on access requirements (white-box vs. black-box), embedding strategy (static vs. dynamic), and the nature of the watermark (zero-bit vs. multi-bit). In the realm of static watermarking, several influential methods have been proposed. Uchida et al. \cite{uchida2017embedding} introduced a pioneering white-box, multi-bit technique where the watermark is embedded directly into the weights of selected convolutional layers through an additional regularization term in the loss function. This approach has laid the foundation for many follow-up works. Building upon this, Li et al. \cite{li2021survey} proposed a Spread-Transform Dither Modulation (ST-DM) based embedding scheme that improves robustness and increases payload capacity without degrading model performance. Tartaglione et al. \cite{tartaglione2021evaluating} developed a zero-bit watermarking strategy that freezes selected weights and enhances their sensitivity, achieving strong robustness to fine-tuning and quantization. Kuribayashi et al. \cite{kuribayashi2019quantization} introduced a quantization-based embedding approach applied via fine-tuning, offering efficient post-training watermarking.

To address concerns over detectability, several works have focused on improving the stealth of watermarks. Wang et al. \cite{wang2019security} proposed embedding watermarks into early-converging weights, while also modeling the network as a graph to improve robustness against structure-preserving attacks. GAN-based approaches have been proposed by Zhang et al. \cite{zhang2020riga}, who introduced the RIGA framework, and further improved by employing adversarial training where a discriminator is used to guide the generator (the model) to embed watermarks in a statistically indistinguishable manner.

Dynamic watermarking methods have also gained traction, particularly for use in black-box settings. These approaches associate the watermark with the behavior of the network when exposed to key inputs. DeepSigns \cite{adi2018turning} is one such method that supports both white-box (activation map-based) and black-box (output-based) watermarking by adding distributional constraints on activations during training. In black-box scenarios, methods such as BlackMarks \cite{yossi2019blackmarks} and Zhang et al. \cite{zhang2018protecting} use specially crafted key inputs (sometimes adversarial or out-of-distribution) that trigger predictable outputs, allowing for verification without internal access. Guo et al. \cite{guo2021invisible} employed differential evolution to optimize invisible trigger locations, enhancing both stealth and robustness. Other approaches have extended watermarking beyond image classification tasks. DeepMarks \cite{chen2019deepmarks} adapted static methods for traitor tracing and was further extended to automatic speech recognition systems  \cite{liu2020adversarial} and hardware-specific watermarking for embedded devices \cite{chen2020hardware}. Similarly, Quan et al. \cite{quan2021robust} and Ong et al. \cite{ong2021watermarking} explored watermarking for generative models, embedding signatures into the output images. Zhao et al.\cite{zhao2021graph} proposed watermarking schemes for Graph Neural Networks (GNNs) using structurally encoded triggers.

Despite significant progress, a major limitation of current schemes lies in the use of handcrafted or static watermark signals, which are susceptible to removal via fine-tuning or smoothing. To mitigate this, researchers have begun to explore the use of chaotic systems for generating watermarks. Chaotic sequences, characterized by high sensitivity to initial conditions and unpredictability, offer a promising direction for enhancing watermark resilience. While methods like that of Guo et al. \cite{guo2021invisible} have utilized evolutionary techniques such as differential evolution to improve watermark robustness, the use of chaotic sequences in combination with heuristic optimization techniques like genetic algorithms remains relatively underexplored.

In this context, our proposed work introduces a novel chaotic sequence-based watermarking framework that embeds the watermark into the weights of a selected layer with minimal perturbation. Verification is performed by extracting the watermark and using a genetic algorithm to optimize the parameters of the chaotic system, ensuring reliable recovery even under post-processing attacks such as fine-tuning. This hybrid approach leverages the unpredictability of chaos and the global search capability of genetic algorithms, contributing to the ongoing development of robust and stealthy DNN watermarking techniques.

\section{Proposed Architecture for DNN Watermarking}
\label{sec:water}

Deep Neural Network (DNN) watermarking is a technique used to protect the intellectual property of trained models by embedding hidden information into the model parameters. This ensures that rightful ownership can be verified even if the model is redistributed or modified. Unlike traditional digital watermarking in images or videos, watermarking a neural network must balance three key goals: robustness (resistance to attacks or modifications), fidelity (preserving model accuracy), and security (preventing unauthorized detection or removal).

The pipeline of our proposed architecture consists of two phases, namely:
\begin{itemize}
    \item \textbf{Watermarking Generation and Embedding Phase}, where a chaotic sequence is generated using predefined parameters (the logistic constant \(r\), the initial value \(x_0\), and a small scaling factor \(\epsilon\), which acts as a secret key) and embedded into the weights of a selected intermediate layer in a pre-trained DNN. The watermarked model is then fine-tuned on the original training data to preserve task accuracy. 
    \item \textbf{Watermarking Verification}, where the embedded watermark is extracted from a suspect DNN and compared with a regenerated chaotic sequence. A genetic algorithm is employed to recover the original chaotic parameters by minimizing the difference between the extracted and regenerated sequences. Successful recovery validates ownership, while failure indicates absence or compromise of the watermark.
\end{itemize}

\begin{figure*}[t] 
    \centering
    \includegraphics[width=\textwidth]{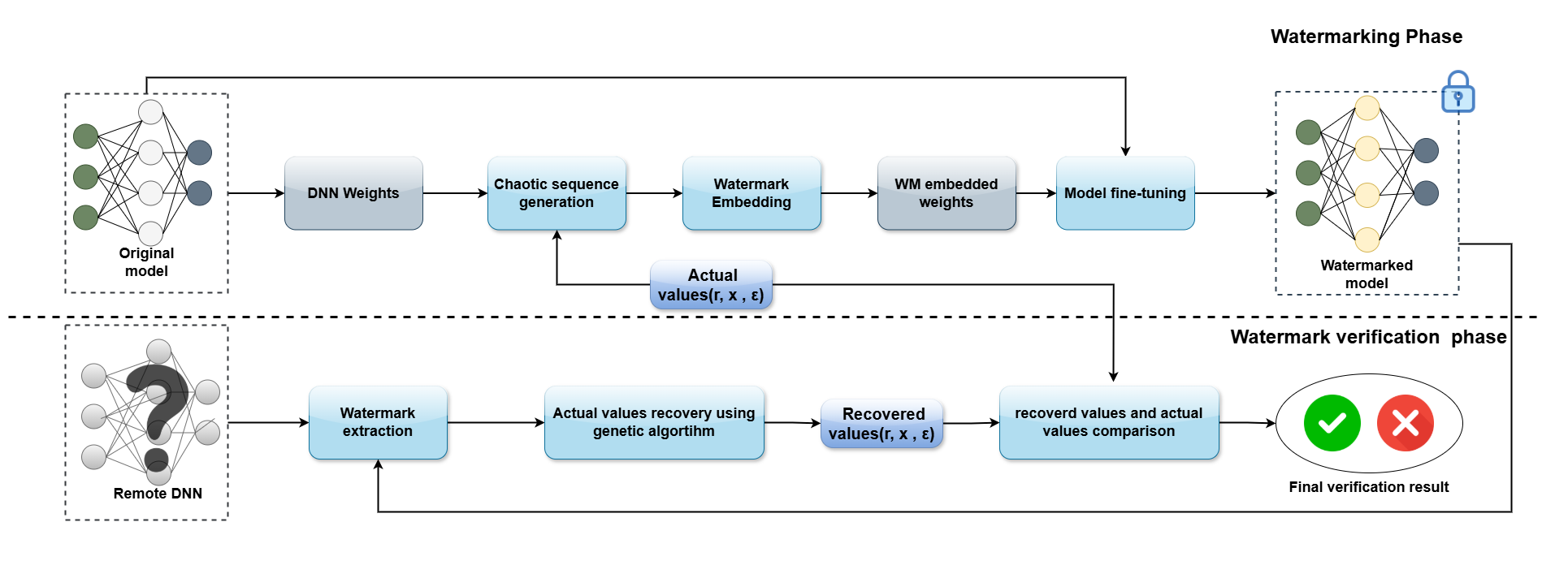}
    \caption{System architecture of the proposed DNN watermarking framework that consists of the Watermarking Generation and Embedding Phase (top) and the Watermark Verification Phase (bottom)}
    \label{fig:architecture}
\end{figure*}

Figure \ref{fig:architecture} illustrates our proposed framework. In the following sections, we detail the two steps of our architecture. Table \ref{tab:SystemSymbols} summarizes the acronyms used in this paper.

\begin{table}
\centering
  \caption{Summary of the acronyms used in the paper}
  \begin{tabular}{ll}
\hline
    \textbf{Symbol} & \textbf{Description}\\
\hline
    DNN & Deep Neural Network\\
    DL & Deep Learning\\
    GA & Genetic Algorithm\\
    IP & Intellectual Property\\
    ML & Machine Learning\\
    MLaaS & Machine Learning-as-a-Service\\
\hline
\end{tabular}
\label{tab:SystemSymbols}
\end{table}

\subsection{Watermark Generation and Embedding}

The watermark generation process begins with the selection of any intermediate layer within the deep neural network. Once the layer is selected, a watermark is generated in the form of a chaotic sequence using the logistic map, which is mathematically defined as:
\begin{equation}
x_{n+1} = r \cdot x_n \cdot (1 - x_n)
\end{equation}

In this equation, \(x_n\)denotes the current value in the sequence, with an initial value \(x_0\) constrained within the interval (0, 1). The parameter \(r\), known as the chaotic control parameter, typically lies within the range of $3.57$ to $4.0$ to ensure chaotic behavior. In our experiment, we selected \(r\) as $3.9$ and \(x_0\) as $0.5$. The logistic map exhibits high sensitivity to initial conditions and parameter values, making it well-suited for watermark generation due to its inherent unpredictability and uniqueness. The length of the chaotic sequence depends on the number of weights present in the selected layer. The chaotic sequence generation function is summarized in Algorithm 1.

\begin{algorithm}
\caption{Generate Chaotic Sequence}
\KwIn{$length \geq 0$, $r = 3.9$, $x = 0.5$}  
\KwOut{Chaotic sequence of given length}  

\If{$length = 0$}{
    \Return empty array\;
}
Initialize $chaotic\_sequence[length]$\;

\For{$i \gets 0$ \KwTo $length - 1$}{
    $x \gets r \cdot x \cdot (1 - x)$\;
    $chaotic\_sequence[i] \gets x$\;
}
\Return $chaotic\_sequence$\;
\end{algorithm}

The chaotic sequence generation function accepts three parameters: the desired length of the sequence, the logistic map parameter \(r\), and the initial value \(x_0\). It begins by checking if the length is zero; if so, it immediately returns an empty list, as no sequence needs to be generated. If the length is valid, it initializes a list of the specified size to store the chaotic values. The function then enters a loop that iteratively applies the logistic map equation to generate each value in the sequence. This process continues until the sequence reaches the required length, resulting in a completely chaotic sequence based on the given parameters.

Once the chaotic sequence is generated, it is embedded into the selected layer by slightly modifying the model's weights. Each weight \(w_i\) in the layer is updated using the following equation:
\begin{equation}
w_i' = w_i + \epsilon \cdot c_i
\end{equation}

\noindent
where \(w_i\) represent the original weight, and 
\(c_i\) denote the corresponding value from the generated chaotic sequence. A small constant \(\epsilon\) is used to control the strength of the watermark.

After embedding the chaotic sequence into the weights of the model. The model with the new weights is then fine-tuned with the original dataset to restore or maintain the model's accuracy. For fine-tuning, the learning rate is reduced by a factor of 10, and the optimizer used is the same as that used for the original training of the model. This fine-tuning step ensures that the watermark is embedded without noticeably affecting the model's performance. After the fine-tuning process, the watermarking is complete.

\subsection{Watermark Verification}

\begin{figure*}[htbp]
\centering
\begin{minipage}[t]{0.48\textwidth}
\centering
\includegraphics[width=\linewidth]{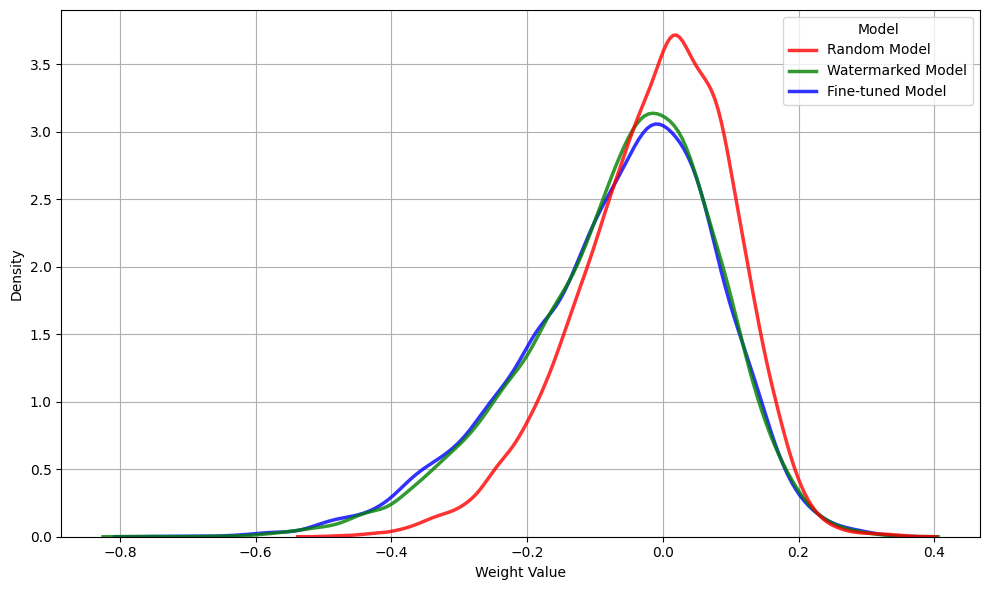}
\vspace{0.5em}
\subcaption{Density Plot of Layer Weights – MNIST Dataset}
\end{minipage}
\hfill
\begin{minipage}[t]{0.48\textwidth}
\centering
\includegraphics[width=\linewidth]{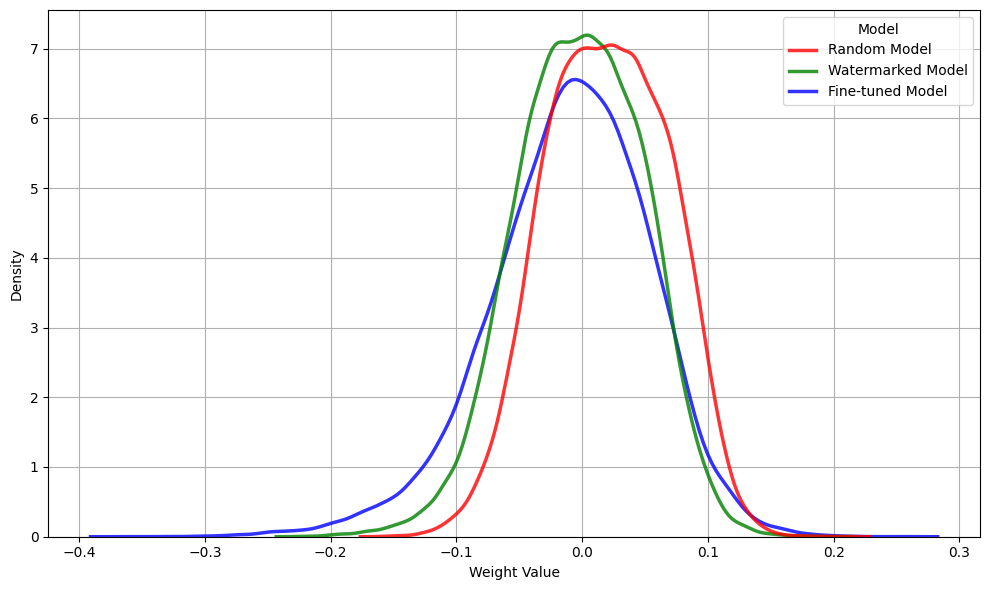}
\vspace{0.5em}
\subcaption{Density Plot of Layer Weights – CIFAR-10 Dataset}
\end{minipage}
\caption{Density plots comparing weight distributions of the Original, Watermarked, and Fine-tuned Models for MNIST and CIFAR-10.}
\label{fig:density-plots}
\end{figure*}

One way to detect the presence of a watermark is by comparing the density of weights in a specific layer across different model instances. In Figure \ref{fig:density-plots}, we illustrate the weight density plots for both the MNIST  and CIFAR-10 datasets. Each subplot compares three models: {\em(i)} a watermarked model, {\em(ii)} a fine-tuned model, and {\em(iii)} a randomly watermarked model using unrelated watermark parameters. 

In both datasets, the weight distribution of the fine-tuned model is closely aligned with the watermarked model, showing only minor variations in peak height and spread. This similarity suggests that the watermark is likely preserved, as fine-tuning has not significantly altered the internal weight structure. On the other hand, the random model shows a noticeably different distribution, indicating a shift caused by embedding unrelated watermark parameters. These observations highlight the usefulness of density analysis as a lightweight, white-box verification method to visually differentiate between models that retain the original watermark and those that do not.

As said before, watermark verification techniques in DNNs are generally classified into two main categories: white-box and black-box approaches. In white-box verification, the model's internal architecture and parameters—such as weights and activations—are accessible to the verifier. This allows for direct inspection of the internal computations, making it possible to detect subtle modifications or hidden patterns introduced during the watermark embedding process. In contrast, black-box verification relies solely on the model's outputs. The verifier provides specific input queries (often called trigger inputs) and observes the corresponding outputs to determine whether the watermark is present. In the proposed watermarking technique, we adopt the white-box verification approach since the watermark is embedded in the intermediate layers of the model. To perform the watermark verification, we implement a genetic algorithm (GA) to estimate the original parameters(\(r\),\(x_0\),\(\epsilon\)) of the logistic map used to generate the watermark. The algorithm is designed to iteratively evolve a population of candidate parameter sets that produce chaotic sequences closely resembling the extracted watermark, measured through both correlation and Mean Squared Error (MSE) metrics.

The first step is the watermark extraction, which is performed by computing the difference between the weights of a specific intermediate layer of the threat model and the watermarked model.  Let \(W_{wm}\)and \(W_{ft}\) represent the weight matrices from the Watermarked and the threat model respectively. The first step involves calculating the difference between these weights to isolate the watermark signal:
\begin{equation}
\Delta W = W_{ft} - W_{wm}
\end{equation}

The difference vector \(\Delta W\) represents the extracted watermark. It is then passed to the genetic algorithm function to evaluate its similarity to the originally embedded chaotic sequence. 

The GA initializes a population of candidate solutions where each individual is a triplet(\(r\), \(x_0\), \(\epsilon\)). A Latin Hypercube Sampling strategy is used for this initialization, ensuring a well-distributed and diverse starting population across the defined parameter ranges. For each candidate solution, the algorithm evaluates how well the corresponding chaotic sequence matches the extracted watermark using a weighted combination of two metrics, namely:

\begin{itemize}
    \item Mean Squared Error (MSE):
    \begin{equation}
\text{MSE}(T, G) = \frac{1}{n} \sum_{i=1}^{n} (T_i - G_i)^2
\end{equation}

    \item Correlation distance:
   \begin{equation}
\text{Corr}(T, G) = 1 - \frac{\sum_{i=1}^{n}(T_i - \bar{T})(G_i - \bar{G})}
{\sqrt{\sum_{i=1}^{n}(T_i - \bar{T})^2} \cdot \sqrt{\sum_{i=1}^{n}(G_i - \bar{G})^2} }
\end{equation}

\end{itemize}

where \(\bar{T}\) and \(\bar{G}\) denote the means of the target and generated sequences respectively. The fitness function combines these metrics linearly:
\begin{equation}
w_{\text{corr}} \cdot \text{Corr}(T, G) + w_{\text{mse}} \cdot \text{MSE}(T, G)
\end{equation}

To evolve the population toward optimal solutions, the genetic algorithm employs a combination of selection, crossover, and mutation strategies. These evolutionary operations are applied iteratively across generations to refine candidate solutions and converge toward the original watermarking parameters.

The selection process utilizes a tournament selection approach, wherein a random subset of individuals is chosen from the current population. Among this group, the individual with the best fitness score based on a combination of mean squared error (MSE) and correlation with the target sequence is selected as a parent. This method maintains diversity while ensuring that higher-quality candidates are more likely to contribute to the next generation.

For generating new individuals, the algorithm applies blend crossover, a recombination technique that creates a child by taking a weighted average of corresponding parameters from two parent solutions. This method enables the algorithm to explore intermediate values in the parameter space, increasing the chances of discovering better-performing solutions. The blending is controlled by a factor \(\alpha\), which adjusts the influence of each parent on the resulting child. The resulting parameter 
\(\text{Child}_i\) for the child is computed using the following equation:
  \begin{equation}
\text{Child}_i = \alpha \cdot \text{Parent}_1^i + (1 - \alpha) \cdot \text{Parent}_2^i \quad \forall i \in \{r, x_0, \epsilon\}
\end{equation}

\noindent 
where \(\text{Parent}_1^i\) and \(\text{Parent}_2^i\) are the \(i^{th}\) parameters from parent 1 and parent 2, respectively, and \(\alpha \in [0,1] \).

To introduce variability and prevent premature convergence, mutation is employed. This operation introduces small Gaussian noise to each parameter (\(r\), \(x_0\), \(\epsilon\)) of the offspring with a certain probability. Mutation ensures continued exploration of the search space, enhances genetic diversity, and helps the algorithm escape from local optima that might otherwise trap the search process.

\begin{algorithm}
\caption{Genetic Algorithm to recover initial values for watermark verification}
\setcounter{AlgoLine}{0}
\KwIn{\textit{target\_sequence, length, pop\_size, generations, r\_range, x0\_range, eps\_range}}
\KwOut{Optimized parameters and MSE}
 \textit{p} $\gets$ \textit{initialize\_population(length, r\_range, x0\_range, e\_range)} \\
 \textit{best\_solution} $\gets$ None \\
\textit{best\_fitness} $\gets \infty$\\ \textit{wait} $\gets 0$,\\ \textit{w\_corr} $\gets 0.03$\\ \textit{w\_mse} $\gets 0.97$\\

\For{\textit{gen} = 0 to \textit{generations}}{
   \textit{fitness\_scores} $\gets$ evaluate\_population(\textit{population}, \textit{target\_sequence}, \textit{length}, \textit{w\_corr}, \textit{w\_mse})\\
   Sort \textit{fitness\_scores} by fitness\\
    \textit{new\_population} $\gets$ top elite individuals\\
    \If{improvement $>$ threshold}{
        Update \textit{best\_solution}, \textit{best\_sequence}, \textit{best\_fitness}\\
     \textit{wait} $\gets 0$
    }\Else{
        \textit{wait} $\gets$ \textit{wait + 1}
    }
    \While{new\_population size $<$ pop\_size}{
        \textit{parent1} $\gets$ tournament\_selection()\\
        \textit{parent2} $\gets$ tournament\_selection()\\
        \textit{child} $\gets$ blend\_crossover(\textit{parent1}, \textit{parent2}, $\alpha$)\\
        \textit{child} $\gets$ mutate(\textit{child}, \textit{r\_range}, \textit{x0\_range}, \textit{e\_range})\\
        Add \textit{child} to \textit{new\_population}\\
    }  \textit{population} $\gets$ \textit{new\_population}\\
}
\textit{r, x0, e} $\gets$ \textit{best\_solution} \\
\Return (\textit{r}, \textit{x0}, \textit{e}, \textit{final\_mse})
\end{algorithm}
An elitism strategy is also implemented, wherein the top-performing individuals are preserved unaltered for the next generation to ensure the retention of the best solutions. This evolutionary process continues for a fixed number of generations or until convergence is observed. The best-performing candidate is then compared against the original parameters. The watermark verification is successful if the difference between the best-performing candidate values and the original values is less. The procedure for the watermark verification is summarized in Algorithm 2.

\section{Experiments}
\label{sec:experiment}

This section presents the experimental evaluation to assess the effectiveness and robustness of our proposed approach. Specifically, we detail the experiment setup, including the employed datasets and metrics, followed by a comprehensive presentation and critical analysis of the obtained results.

\subsection{Experiments setup}
The datasets used to train the two models for our experiments, which are directly imported using the Keras API, are the following:

\begin{itemize}
    \item \textbf{MNIST} (Modified National Institute of Standards and Technology) dataset, is an established standard for handwritten digit recognition applications. The collection includes $70,000$ grayscale images with dimensions of 28×28 pixels and ten numbers (0–9). The data collection includes $60,000$ training images and $10,000$ test images. The MNIST dataset serves as a foundation for deep learning models by making image classification algorithm testing easier. MNIST provides foundational support for beginning with sophisticated image recognition tasks, and researchers typically use it to assess several machine learning algorithms in conjunction with CNNs.
    \item \textbf{CIFAR-10} (Canadian Institute for Advanced Research - 10) dataset, which is commonly used for image classification tasks. The CIFAR-10 collection includes $60,000$ color photos with a resolution of 32×32 pixels, divided into 10 categories (e.g., airplanes, autos, birds, cats, deer, dogs, frogs, horses, ships, and trucks). The dataset is divided into $50,000$ training and $10,000$ test images. CIFAR-10 is suitable for testing advanced deep learning models and convolutional neural networks. The issue in classification derives from the small image size paired with the wide range of object categories that have made deep learning and computer vision research popular with CNNs.
\end{itemize}

To prepare the data for training, normalization is applied to scale the pixel values from their original range of 0 to 255 down to a standard range of 0 to 1. This is done by converting the integer pixel values into floating-point numbers and dividing each by 255. Normalization helps the neural network train more efficiently by improving numerical stability, speeding up convergence, and allowing the model to learn better from the input features. In addition to pre-processing the input data, the class labels are also converted into a machine-readable format using label encoding and one-hot encoding. Label encoding assigns a unique numeric value to each class, and then one-hot encoding transforms these numeric labels into binary vectors, where only the index of the true class is marked as 1 and the rest are set to 0. After the preprocessing stage, two separate deep learning models were trained, one on the CIFAR-10 dataset and the other on the MNIST dataset, to test our watermarking method.

To assess the performance of our defense strategy, we mainly employed the following evaluation metrics:

\begin{itemize}
    \item \textbf{Accuracy}, that quantifies the proportion of correctly classified instances among the total number of samples. It is formally defined as:
    $$\text{A} = \frac{1}{N} \sum_{i=1}^{N} (y_i = \hat{y}_i)$$
    where $N$ is the total number of instances, $y_i$ is the true label for the $i$-th instance, and $\hat{y}_i$ is the predicted label for the $i$-th instance.

     \item \textbf{Recall}, which measures the proportion of correctly classified samples among all samples that truly belong to that class.
     $$\text{R}_s = \frac{\text{TP}_s}{\text{TP}_s + \text{FN}_s}$$
     where TP (True Positives) refers to the number of instances correctly predicted as belonging to a particular class, and FN (False Negatives) refers to the number of instances that belong to a class but were not predicted as such.

     \item \textbf{F1-score}, which is the harmonic mean of precision and recall.
     $$\text{F1-score} = 2 \cdot \frac{\text{P} \cdot \text{R}}{\text{P} + \text{R}}$$
     where Precision (P) is defined as the proportion of correctly predicted labels (TP) among all predicted labels (TP + FP), $\text{P} = \frac{\text{TP}}{\text{TP} + \text{FP}}$.

\end{itemize}

\subsection{Fidelity evaluation}

To ensure that the proposed watermarking technique does not compromise the model's ability to perform its original task, we evaluate the fidelity of the watermarked model by comparing its accuracy with that of the original (non-watermarked) model. Fidelity in this context refers to the degree to which the model's predictive performance is preserved after the watermark has been embedded. To evaluate the fidelity of our proposed watermarking approach, the models of both datasets are tested on the same testing dataset used for testing the original training of the model. On the MNIST dataset, we observed a marginal improvement in accuracy after watermarking. This slight gain suggests that the watermarking process may have introduced a regularizing effect, potentially aiding generalization. In contrast, for the CIFAR-10 dataset, a minor drop in accuracy was noted. However, the decline was minimal, and the overall performance remained within acceptable limits.

These results demonstrate that the proposed watermarking method maintains the predictive capabilities of the original models. The watermark does not introduce any disruptive changes to the model's behavior, highlighting the fidelity of the technique. Tables \ref{tab:sidebyside} and \ref{tab:sidebyside_below} show how the models perform before and after watermarking in both CIFAR-10 and MNIST datasets.

\begin{table*}[htbp]
\centering
\caption{Performance of the model before and after watermarking (CIFAR-10)}
\begin{minipage}[t]{0.48\textwidth}
\centering
\begin{tabular}{c|c|c|c}
\textbf{Class} & \textbf{Precision} & \textbf{Recall} & \textbf{F1}   \\
\hline
Airplane & 0.8316 & 0.7260 & 0.7752 \\
Automobile & 0.8598 & 0.8890 & 0.8741 \\
Bird & 0.6762 & 0.5700 & 0.6186 \\
Cat & 0.4898 & 0.6240 & 0.5488  \\
Deer & 0.6143 & 0.7550 & 0.6774 \\
Dog & 0.6891 & 0.5320 & 0.6005 \\
Frog & 0.6699 & 0.8910 & 0.7648 \\
Horse & 0.8756 & 0.7110 & 0.7848 \\
Ship & 0.8770 & 0.8270 & 0.8513 \\
Truck & 0.9000 & 0.8010 & 0.8476  \\
\hline
\textbf{Acc.} & - & - & \textbf{0.7326} \\
\textbf{Macro} & 0.7483 & 0.7326 & 0.7343 \\
\textbf{Weighted} & 0.7483 & 0.7326 & 0.7343 \\
\end{tabular}
\vspace{0.5em}
\subcaption{Original Model (Before Watermarking)}
\end{minipage}
\hfill
\begin{minipage}[t]{0.48\textwidth}
\centering
\begin{tabular}{c|c|c|c}
\textbf{Class} & \textbf{Precision} & \textbf{Recall} & \textbf{F1}  \\
\hline
Airplane & 0.8438 & 0.6860 & 0.7568 \\
Automobile & 0.9013 & 0.8490 & 0.8744 \\
Bird & 0.6289 & 0.6100 & 0.6193 \\
Cat & 0.4967 & 0.5960 & 0.5418 \\
Deer & 0.5948 & 0.8030 & 0.6834 \\
Dog & 0.6880 & 0.5600 & 0.6174 \\
Frog & 0.6553 & 0.9300 & 0.7595 \\
Horse & 0.9123 & 0.6660 & 0.7699  \\
Ship & 0.8734 & 0.8350 & 0.8538 \\
Truck & 0.9103 & 0.7710 & 0.8349 \\
\hline
\textbf{Acc.} & - & - & \textbf{0.7279} \\
\textbf{Macro} & 0.7505 & 0.7279 & 0.7311  \\
\textbf{Weighted} & 0.7505 & 0.7279 & 0.7311 \\
\end{tabular}
\vspace{0.5em}
\subcaption{Watermarked Model}
\end{minipage}
\label{tab:sidebyside}
\end{table*}

\begin{table*}[htbp]
\centering
\caption{Performance of the model before and after watermarking (MNIST)}
\begin{minipage}[t]{0.48\textwidth}
\centering
\begin{tabular}{c|c|c|c}
\textbf{Class} & \textbf{Precision} & \textbf{Recall} & \textbf{F1} \\
\hline
0 & 0.9959 & 0.9969 & 0.9964 \\
1 & 0.9947 & 0.9991 & 0.9969 \\
2 & 0.9952 & 0.9952 & 0.9952 \\
3 & 0.9941 & 0.9950 & 0.9946 \\
4 & 0.9939 & 0.9929 & 0.9934 \\
5 & 0.9944 & 0.9333 & 0.9938  \\
6 & 0.9948 & 0.9948 & 0.9948 \\
7 & 0.9932 & 0.9942 & 0.9937  \\
8 & 0.9969 & 0.9949 & 0.9959 \\
9 & 0.9920 & 0.9881 & 0.9901 \\
\hline
\textbf{Acc.} & - & - & \textbf{0.9945} \\
\textbf{Macro} & 0.9945 & 0.9944 & 0.9945  \\
\textbf{Weighted} & 0.9945 & 0.9945 & 0.9945 \\
\end{tabular}
\vspace{0.5em}
\subcaption{Original Model (Before Watermarking)}
\end{minipage}
\hfill
\begin{minipage}[t]{0.48\textwidth}
\centering
\begin{tabular}{c|c|c|c}
\textbf{Class} & \textbf{Precision} & \textbf{Recall} & \textbf{F1} \\
\hline
0 & 0.9959 & 0.9980 & 0.9969 \\
1 & 0.9930 & 1.0000 & 0.9965 \\
2 & 1.0000 & 0.9952 & 0.9976 \\
3 & 0.9980 & 0.9960 & 0.9970 \\
4 & 0.9949 & 0.9959 & 0.9954 \\
5 & 0.9944 & 0.9955 & 0.9950 \\
6 & 0.9958 & 0.9916 & 0.9937 \\
7 & 0.9942 & 0.9961 & 0.9951 \\
8 & 0.9959 & 0.9969 & 0.9964 \\
9 & 0.9950 & 0.9911 & 0.9930 \\
\hline
\textbf{Acc.} & - & - & \textbf{0.9957} \\
\textbf{Macro} & 0.9957 & 0.9956 & 0.9957 \\
\textbf{Weighted} & 0.9957 & 0.9957 & 0.9957 \\
\end{tabular}
\vspace{0.5em}
\subcaption{Watermarked Model}
\end{minipage}
\label{tab:sidebyside_below}
\end{table*}

\subsection{Watermark Detection using  Logistic regression classifiers}
To detect whether a DNN model has been watermarked or tampered with (such as fine-tuned), we designed a lightweight post-hoc classifier based on logistic regression. This classifier uses the internal activation patterns of the DNN to differentiate between original, watermarked, and fine-tuned models. 
The core idea is that watermarking introduces subtle but consistent changes in the weight distribution of a DNN, which affects the intermediate activations when the model processes inputs. Similarly, fine-tuning may also alter the model’s internal behavior. These changes are often not visible at the output layer but can be captured in the activation values of hidden layers.

To utilize this, we first selected a specific layer (or Watermarked layer) and extracted the activation outputs for a fixed set of input samples. For each input image passed through the model, we recorded the activations of the chosen layer and flattened them to form a feature vector.

To ensure that only confident predictions are used, thereby reducing noise from misclassified or uncertain inputs, we filtered the samples based on their softmax prediction scores. For the MNIST dataset, only samples with prediction confidence above $0.9$ were considered. For CIFAR-10, a slightly lower threshold of $0.7$ was used due to the more complex nature of the dataset. These thresholds were chosen empirically to balance between data quality and sample availability.

Training the Logistic Regression Classifier
Once the feature vectors were collected, we labeled them according to the source model: original, watermarked, or fine-tuned. These labeled activation features formed the training dataset for our logistic regression model. We trained the classifier using standard techniques, minimizing the cross-entropy loss and optimizing the weights to separate the three classes.

Logistic regression was chosen because of its simplicity and interpretability, as well as its ability to generalize well on high-dimensional but linearly separable activation data. Despite being a shallow model, it performed exceptionally well in distinguishing the subtle differences among the model variants.

After training, the classifier was tested on activation data from unseen images, again filtered using the same prediction score thresholds. The performance of the classifier is illustrated using the confusion matrices in Figure 3 for both the MNIST and CIFAR-10 datasets.

\begin{figure*}[htbp]
\centering
\begin{minipage}[t]{0.48\textwidth}
\centering
\includegraphics[width=\linewidth]{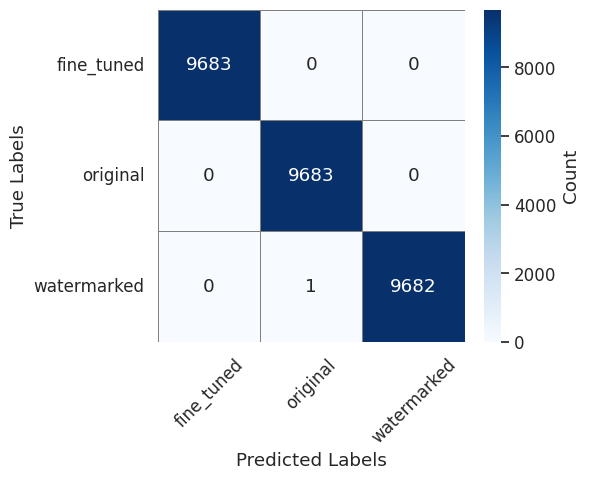}
\vspace{0.5em}
\subcaption{Confusion Matrix – MNIST Dataset}
\end{minipage}
\hfill
\begin{minipage}[t]{0.48\textwidth}
\centering
\includegraphics[width=\linewidth]{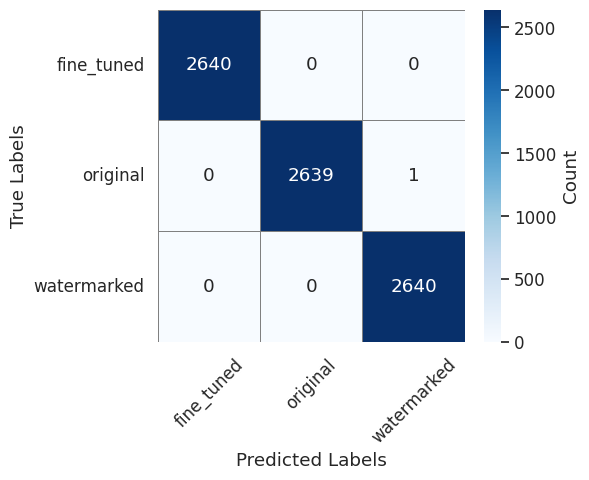}
\vspace{0.5em}
\subcaption{Confusion Matrix – CIFAR-10 Dataset}
\end{minipage}
\caption{Confusion matrices showing classification performance of the logistic regression classifiers. The classifiers identify the model based on its activation features.}
\label{fig:confusion-matrices}
\end{figure*}

\begin{itemize}
    \item For MNIST, the classifier achieved near-perfect classification with only a single misclassification out of $7,920$ samples.
    \item For CIFAR-10, which is more challenging, the classifier still performed with high accuracy, making only one error out of $29,049$ samples.
\end{itemize}

These results strongly indicate that the internal activations of a DNN carry distinguishable patterns depending on whether the model is original, watermarked, or fine-tuned. The effectiveness of the logistic regression classifier confirms that activation-based features are a reliable signal for post-hoc watermark verification.

\subsection{Effectiveness}

To evaluate how effective the proposed watermarking and verification approach is, we conducted two experiments to simulate both ownership confirmation and resistance to false claims. The core idea was {\em(i)} to test whether our genetic algorithm–based verification process could reliably retrieve the original chaotic parameters (\(r\), \(x_0\), \(\epsilon\)) from a model that contains the watermark and {\em(ii)} ensure it does not falsely recognize watermark ownership in models that were not embedded with our specific sequence.

\begin{enumerate}
    \item \textbf{Ownership Confirmation}. We considered a model that was previously watermarked using our chaotic sequence embedding technique. This model underwent the watermark verification process using the genetic algorithm. The outcome was highly successful; the recovered parameters matched the original values used during watermark embedding with a very small tolerance, thus confirming the watermark's presence. This clearly demonstrates that our method can reliably verify ownership.
    \item \textbf{Resistance to False Claims}. We tested the same watermark verification process on a randomly watermarked model. This model had been embedded with a chaotic sequence generated using different random parameters(\(r\), \(x_0\), \(\epsilon\)) that were unrelated to our original watermark. When the genetic algorithm attempted to recover the original watermark parameters from this model, it failed to find a close match. The recovered values were significantly different from the expected ones, and the mean square error remained high, indicating a poor match. This result is important because it shows that our method is not only effective in confirming true ownership but also robust against false positives; it does not mistakenly verify ownership in unrelated models.
\end{enumerate}

Together, these experiments highlight the strong effectiveness of our watermarking system. It can accurately confirm the presence of the watermark when it exists and reliably reject unrelated models that do not carry the original watermark. This dual capability ensures that the verification process is both precise and secure, making it a practical solution for real-world intellectual property protection of deep neural network models. Table \ref{tab:ga_verification_blocked} clearly demonstrates the effectiveness of the proposed watermark verification method using a genetic algorithm. When the algorithm is applied to the watermarked models, it successfully recovers parameter values that are very close to the actual values used during the watermark embedding process. For both MNIST and CIFAR-10 datasets, the recovered values of \(r\), \(x_0\), \(\epsilon\) show minimal differences from the original parameters, with deviations as small as 0.001 for some values. In contrast, when the same verification process is applied to a random model that does not contain the watermark, the recovered parameters differ significantly from the actual values.

\begin{table*}[htbp]
\centering
\caption{Watermark Verification Results Using Genetic Algorithm (fine-tuned model vs. Random Model)}
\begin{tabular}{|c|c|p{3.5cm}|c|p{3.5cm}|c|}
\hline
\textbf{Parameter} & \textbf{Actual Value (A)} & \textbf{Recovered Values from attacked model (B)} & \textbf{Diff (|A - B|)} & \textbf{ Recovered values from a Random model (C)} & \textbf{Diff (|A - C|)} \\
\hline
\multicolumn{6}{|c|}{\textbf{MNIST}} \\
\hline
$r$        & 3.900000 & 3.911288 & 0.011288 &3.646363 
& 0.253637 \\
$x_0$      & 0.500000 & 0.497336 & 0.002664 & 0.270680  & 0.229320 \\
$\epsilon$ & 0.010000 & 0.011182 & 0.001182 &0.050000 
& 0.040000\\
\hline
\multicolumn{6}{|c|}{\textbf{CIFAR-10}} \\

\hline
$r$        & 3.900000 & 3.912737 & 0.012737 & 3.600858 & 0.299142\\
$x_0$      & 0.500000 & 0.504805 & 0.004805 & 0.375127 & 0.124873\\
$\epsilon$ & 0.010000 & 0.010802 & 0.000802 &0.035241& 0.025241\\
\hline
\end{tabular}
\label{tab:ga_verification_blocked}
\end{table*}

\begin{figure*}
\centering
\begin{minipage}[t]{0.48\textwidth}
\centering
\includegraphics[width=\linewidth]{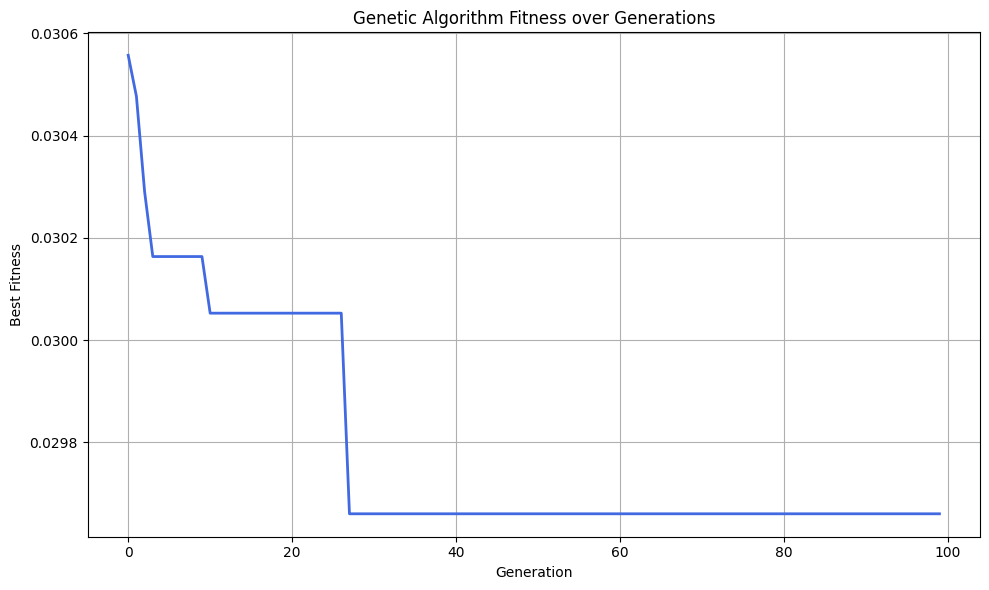}
\vspace{0.5em}
\subcaption{Fitness Convergence – MNIST Dataset}
\end{minipage}
\hfill
\begin{minipage}[t]{0.48\textwidth}
\centering
\includegraphics[width=\linewidth]{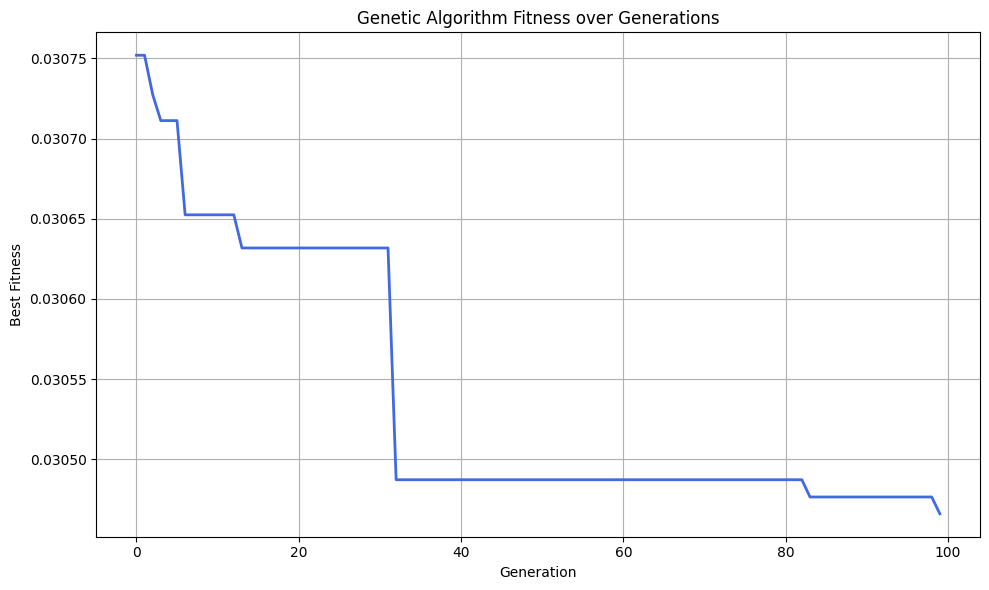}
\vspace{0.5em}
\subcaption{Fitness Convergence – CIFAR-10 Dataset}
\end{minipage}
\caption{Best fitness value of the Genetic Algorithm over generations for watermark recovery from fine-tune attacked models on MNIST and CIFAR-10 datasets.}
\label{fig:fitness-convergence}
\end{figure*}

\begin{figure*}[htbp]
\centering
\begin{minipage}[t]{0.48\textwidth}
\centering
\includegraphics[width=\linewidth]{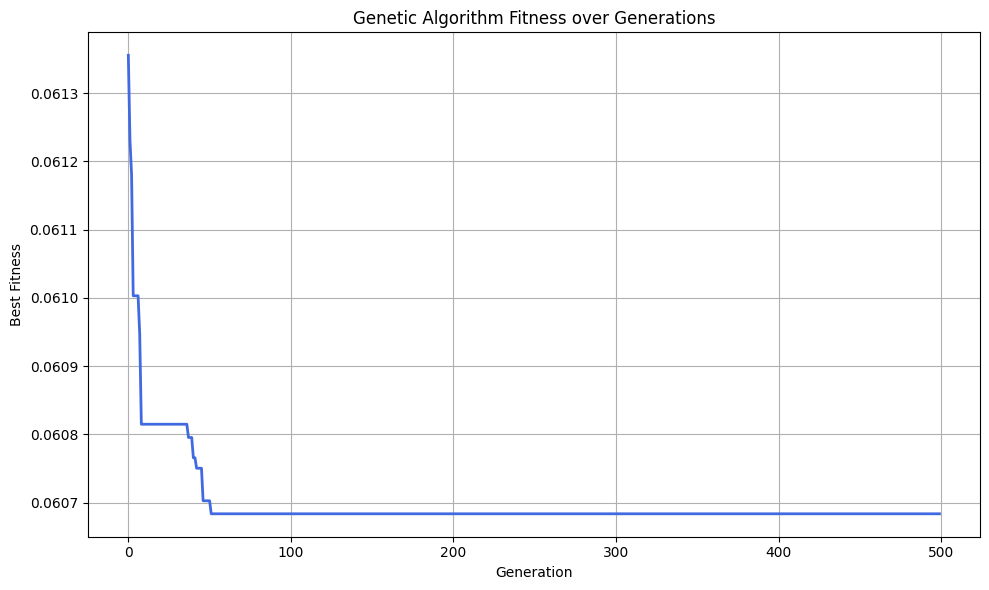}
\vspace{0.5em}
\subcaption{Fitness Convergence – MNIST Dataset}
\end{minipage}
\hfill
\begin{minipage}[t]{0.48\textwidth}
\centering
\includegraphics[width=\linewidth]{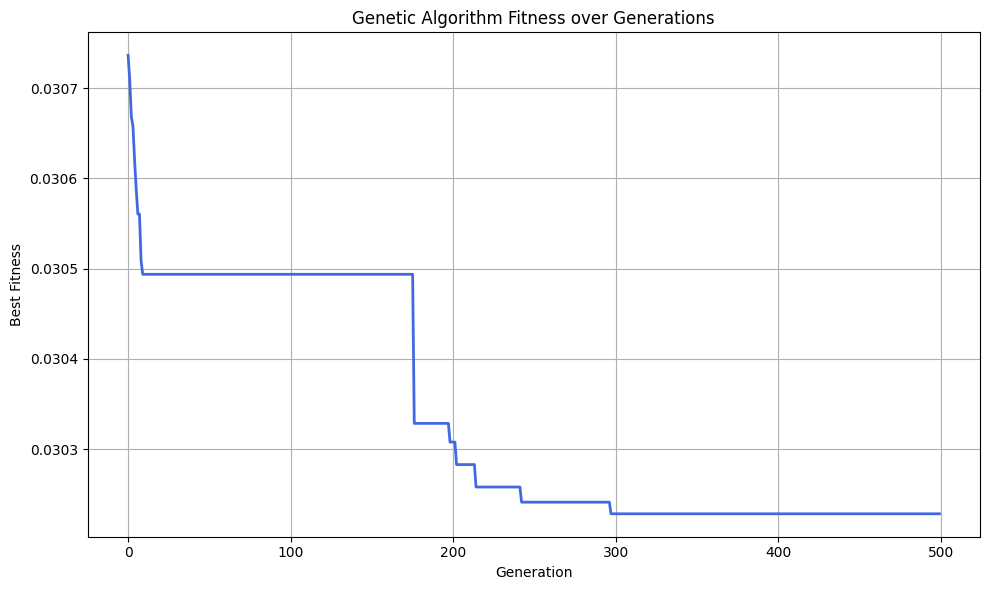}
\vspace{0.5em}
\subcaption{Fitness Convergence – CIFAR-10 Dataset}
\end{minipage}
\caption{Best fitness value of the Genetic Algorithm over generations for watermark recovery from random models on MNIST and CIFAR-10 datasets.}
\label{fig:fitness-convergence2}
\end{figure*}

Also, the convergence plots in Figures \ref{fig:fitness-convergence} and \ref{fig:fitness-convergence2} demonstrate that the genetic algorithm consistently converges to an optimal solution in both cases when verifying a model that contains the watermark and when applied to a random, non-watermarked model. This behavior confirms that the genetic algorithm performs a stable and deterministic search rather than producing arbitrary or random outputs. Although the final convergence error is significantly lower for the watermarked model, the consistent convergence across both scenarios validates the robustness and reliability of the optimization process used in watermark verification. This disparity confirms that the genetic algorithm is capable of accurately identifying the presence of the watermark by retrieving the correct chaotic parameters only from the models that have been watermarked. Thus, the results of the reliability of the proposed watermark verification method.

\subsection{Robustness}

The final experiment aims to test how strong and reliable our proposed watermarking method is. To do so, we performed an attack known as a fine-tuning attack. A fine-tuning attack is a common technique used by adversaries to subtly change a pre-trained model by continuing its training on a similar dataset. The goal of this attack is to slightly adjust the model's parameters in hopes of damaging or completely removing the embedded watermark, while still keeping the model’s performance intact. This type of attack is effective in white-box scenarios, where the attacker has full knowledge of the model's structure, training data, and other hyperparameters. In our setup, we assumed the attacker had complete access to the watermarked model, including its architecture and training configuration. To simulate the attack, we fine-tuned the model using the same dataset it was originally trained on. Specifically, we took the test dataset used during watermark verification and split it in half, using the first half for fine-tuning and the second half for evaluating the modified model. This ensured the attack used realistic data without introducing external variations. For fine-tuning, we used the same optimizer as in the original training, but with a learning rate reduced by 10 times to mimic subtle parameter shifts. These gradual changes reflect how attackers would typically modify a model in practice, aiming to bypass detection mechanisms without harming model performance. After completing the fine-tuning, we tested whether the watermark could still be identified. We extracted the weights from the originally watermarked layer and applied our genetic algorithm to search for the chaotic sequence parameters used during embedding. Despite the model undergoing several rounds of fine-tuning, we were still able to successfully detect and verify the watermark. This confirms that the watermark remains intact even after attempts to tamper with the model. Overall, our results demonstrate the robustness of the proposed watermarking technique against fine-tuning attacks, highlighting its ability to withstand realistic and informed adversarial scenarios. 

The results presented in Table \ref{tab:ga_verification_blocked} further highlight the robustness of the watermark verification method against fine-tuning attacks. The recovered parameter values for both MNIST and CIFAR-10 datasets, obtained from fine-tuned models that originally contained the watermark, remain closely aligned with the actual watermarking parameters. Despite the fine-tuning process, which typically alters the model weights and may attempt to remove or weaken the watermark, the genetic algorithm is still able to accurately recover values for \(r\), \(x_0\), \(\epsilon\) with minimal deviations. For example, in the MNIST case, the differences between actual and recovered values remain within a very small margin. This demonstrates that the watermark is not only detectable but also resilient to model modifications such as fine-tuning, reinforcing the effectiveness of the watermarking and verification approach in real-world adversarial scenarios.

Figure \ref{fig:fitness-convergence} illustrates the fitness convergence curves of the genetic algorithm during the process of recovering the actual watermarking parameters from fine-tuned models. In both cases (MNIST and CIFAR-10), the fitness value, which represents the error between the recovered chaotic sequence and the extracted difference sequence, consistently decreases over generations, ultimately stabilizing at a low value. This steady decline indicates that the genetic algorithm is effectively optimizing the parameters and successfully converging towards the true values embedded in the model, despite the presence of fine-tuning attacks. The convergence to a minimal fitness value validates the algorithm’s ability to accurately recover watermark information, demonstrating its robustness even under adversarial conditions.

\section{Discussion and Limitations}
\label{sec:discussion}

While the proposed watermarking method achieves strong fidelity and robustness, it has several limitations that are important to consider. First, the method is designed for a white-box setting, where the internal weights of the model are accessible to the verifier. This limits its application in black-box scenarios such as commercial APIs or encrypted inference platforms, where only output access is available. 

Another limitation is the dependency on model architecture. Although our technique is theoretically applicable to any DNN, all experiments in this study were conducted on convolutional neural networks (CNNs) for image classification tasks. The behavior of chaotic watermark embedding in other architectures like transformers, recurrent neural networks, or language models remains unexplored and should be addressed in future work. The watermark verification step involves running a genetic algorithm to recover the original chaotic parameters. While effective, this process can be computationally expensive, especially for large models or longer watermark sequences. The runtime is acceptable for small-scale datasets such as MNIST and CIFAR-10, but may become impractical in real-time or large-scale deployments without optimization.

Additionally, although our watermark shows resilience against fine-tuning, we have not evaluated its robustness against other common attacks such as pruning and watermark overwriting. A comprehensive evaluation under these threat models would further strengthen confidence in its real-world applicability. The watermarking mechanism depends on three parameters: \(r\), \(x_0\), and \(\epsilon\), used to generate a chaotic sequence. Selecting these values requires care; if \(\epsilon\) is too small, the watermark becomes fragile and hard to recover; if it is too large, it may degrade model accuracy or become detectable. Similarly, certain combinations of \(r\) and \(x_0\) might lead to non-robust sequences. Unlike some prior works that support multi-bit watermarking to embed identifiers or metadata, our approach only embeds a continuous chaotic sequence without direct support for encoding binary payloads. This may limit its usefulness in scenarios requiring large or structured ownership information.

Finally, the success of watermark verification depends on the threshold values used to determine parameter recovery accuracy. Improper threshold calibration could lead to either false positives, wrongly confirming ownership, or false negatives, failing to verify the watermark in a genuinely marked model.

Despite these limitations, the proposed approach serves as an efficient white-box watermarking method that enables watermark embedding without requiring retraining of the model. It maintains a balance between subtle watermark integration and reliable parameter recovery, making it suitable for scenarios where internal model access is available and lightweight post-training watermarking is needed.

\section{Conclusion}
\label{sec:conclusion}

In this work, we presented a white-box watermarking framework for DNNs that leverages the properties of chaotic sequences for ownership embedding and genetic algorithms for verification. The watermark is embedded by injecting a carefully crafted chaotic sequence into the weights of a selected intermediate layer in a trained model. This process is followed by a brief fine-tuning step to preserve the model’s performance. Unlike many prior watermarking schemes, our approach does not modify the training procedure or require access to original training data, making it suitable for post-training watermarking of already deployed models. The verification process is based on recovering the original chaotic parameters used in watermark generation. By computing the difference between a reference watermarked model and a potentially modified model, we isolate the watermark signal and use a genetic algorithm to search for the parameters that regenerate the closest matching chaotic sequence. Our experiments on both MNIST and CIFAR-10 datasets show that the embedded watermark remains detectable even after fine-tuning attacks, demonstrating robustness under realistic post-deployment modifications. In addition to quantitative evaluations, we used activation-based logistic regression classifiers and weight density plots to support the presence of the watermark and to visually distinguish watermarked models from non-watermarked and tampered ones. These auxiliary detection techniques complement the parameter recovery process and provide lightweight verification tools when full-scale recovery is unnecessary. While the method performs well in controlled settings, several limitations were identified. The approach assumes full white-box access to the model's internal weights and has been tested primarily on CNN architectures for image classification tasks. The reliance on parameter recovery via evolutionary search also introduces additional computation time during verification. Furthermore, the method currently supports only single-sequence watermarking and does not directly encode multi-bit identifiers.

Future work can explore several directions to enhance the generalizability and robustness of the proposed watermarking method. One possible extension is to adapt the framework for a broader range of model architectures, including transformers, recurrent neural networks, and large-scale language models. Improving verification efficiency through faster search strategies or the use of lightweight surrogate models for parameter recovery is another promising direction. Additionally, enabling support for multi-bit watermarking would allow for the embedding of more complex ownership metadata, such as creator identifiers, licensing terms, or version control information. Further investigations could evaluate the resilience of the watermark against a wider set of adversarial attacks, including weight pruning, quantization, adversarial retraining, and model extraction. Developing hybrid approaches that combine structural watermarking with behavioral signals may also help in building more robust verification techniques. Finally, extending the current method to support black-box or semi-black-box scenarios could significantly increase its applicability in real-world deployment settings where internal model access is limited.

Overall, this work contributes a practical and flexible watermarking approach that adds to the growing toolkit of techniques for securing deep neural networks against unauthorized use and redistribution.

\bibliographystyle{plain}
{\footnotesize
\bibliography{biblio}
}
\end{document}